\documentclass[12pt]{revtex4}
\usepackage[dvips]{graphicx}
\usepackage{delarray}
\usepackage{amsmath}
\usepackage{amssymb}
\usepackage{here}
\usepackage{color}

\newcommand{\vt}{\vartheta}
\newcommand{\ord}{\mathcal{O}}
\newcommand{\be}{\begin{equation}}
\newcommand{\ee}{\end{equation}}
\newcommand{\ps}{Pfirsch-Schl\"{u}ter}
\newcommand{\lla}{\left\langle}
\newcommand{\rra}{\right\rangle}

\newcommand{\vpa}{v_{\|}}

\newcommand{\xpa}{x_{\|}}
\newcommand{\xpe}{x_{\perp}}

 \newcommand{\g}[1]{\mbox{\boldmath $#1$}}

 \newcommand{\lp}{\left(}
\newcommand{\rp}{\right)}

\begin{document}

\begin{center}
\Large

{\bf Impurity flows and plateau-regime poloidal density variation in a tokamak pedestal}\\
~\\*[0.5cm] \normalsize
M. Landreman$^{1}$, T. F\"ul\"op$^{2}$, D Guszejnov$^{3}$\\
{\it\small
$^{1}$ Plasma Science and Fusion Center, MIT, Cambridge, MA, 02139, USA\\
$^{2}$ Department of Applied Physics, Nuclear Engineering, Chalmers
  University of Technology and Euratom-VR Association, G\"oteborg,
  Sweden\\
$^{3}$ Department of Nuclear Techniques, Budapest University of Technology and Economics, Association EURATOM, H-1111 Budapest, Hungary}\\
\today
\end{center}
\begin{abstract}
  In the pedestal of a tokamak, the sharp radial gradients of density
  and temperature can give rise to poloidal variation in the density
  of impurities.  At the same time, the flow of the impurity species
  is modified relative to the conventional neoclassical result.  In
  this paper, these changes to the density and flow of a collisional
  impurity species are calculated for the case when the main ions are
  in the plateau regime.
  In this regime it is found that the impurity density can be higher at either
  the inboard or outboard side.
  This finding differs from earlier results for banana- or \ps-regime main ions,
  in which case the impurity density is always higher at the inboard side in the absence of rotation.
  Finally, the modifications to the impurity flow are
  also given for the other regimes of main-ion collisionality.
\end{abstract}
\maketitle
\section{Introduction}
Conventional theory of neoclassical transport in tokamaks \cite{HH,HS}
is not applicable to regions where the pressure and temperature
profiles are very steep, such as the pedestal at the plasma edge.  As
the radial scale length decreases, poloidal variation arises in the
temperature and density of each species.  In an impure plasma,
typically the first quantity to develop a poloidal variation is the
impurity density, and indeed, strong poloidal impurity asymmetries
have been observed in experiments \cite{marr, newcmod}.
In
Refs.~\cite{perbif,fh1,fh2}, neoclassical theory for an impure plasma
was extended to allow for larger gradients than are usually
considered. Specifically, the gradients were allowed to be so large
that the friction between the bulk ions and heavy impurity ions could
compete with the parallel impurity pressure gradient, as is typically
the case in the tokamak edge.  Mathematically, this means that the
parameter
$\Delta \equiv \delta \hat{\nu}_{ii} z^2 $ was assumed to be of order
unity, but the poloidal Larmor radius of the bulk ions divided by the
radial scale length associated with the density and temperature
profiles $ \delta = \rho_\theta/L_{\perp}$ was assumed to be small. Here
$z$ is the impurity charge number,
$\hat{\nu}_{ii}=L_\parallel/\lambda_i$ is a measure of the ion collisionality,
$\lambda_i$ is the bulk ion mean-free path, and $L_\parallel$ is the
connection length. It was shown that the impurity dynamics then
become nonlinear, and if the pressure and temperature gradients of the
main ion species are sufficiently steep, the impurities are pushed to the
inboard side of the flux surface.

Recently, the in-out density asymmetry $A = n_H / n_L$ was measured for boron
impurities in Alcator C-Mod ~\cite{marr}.  Here, $n_H$ and $n_L$ refer respectively
to the impurity density
at the high-field-side midplane and low-field-side midplane of a given flux surface.
It was observed that $A$ could be either less than or greater than one.
A comparison was made to a theoretical model of impurity asymmetry in strong gradient regions~\cite{fh2}
in which the primary ion species was assumed to
be in the \ps\ regime of collisionality.  This model predicts that $A$ must be more than one, and for
the parameters of the Alcator C-Mod experiments, the predicted $A$ was systematically
closer to unity than the measured ratio.
One factor which likely contributes to the discrepancy is
that much of the data were taken in a region in which the main ions were in the plateau collisionality regime
rather than the \ps\ regime.
Reference~\cite{marr} therefore suggests that an analogous theoretical model should be developed
for the plateau regime, and it is the purpose of this paper to present such a model.
Impurity asymmetry in the banana collisionality regime has been analyzed
previously in~\cite{perbif,fh1}.
Other than the collisionality, the present work
uses the same orderings as the previous models: $\Delta \sim 1, z \gg 1,$ and $\delta \ll 1$.

The poloidal rearrangement of the impurities
affects the impurity velocity due to the requirement of mass conservation.
In previous work on the banana and \ps\ regimes,
this alteration to the impurity flow was not explicitly calculated.
However, pedestal impurity flows are measured routinely in experiments
~\cite{marr,newcmod}, so impurity flows represent
an important point of comparison between experiment and theory.
The measurements and conventional neoclassical theory often disagree.
In particular, when the main ions are in the plateau or banana collisionality
regime, the measured impurity flow is greater in the direction of the electron diamagnetic velocity
than predicted.
Consequently, in this paper we give explicit forms
for the modified impurity flows, and we examine whether
the modifications are sufficient to reconcile neoclassical
theory with the experimental measurements.

The remainder of the paper is organized as follows. In
Sec.~\ref{sec:ions} we describe the kinetics of main ions in the
plateau regime.  In Sec.~\ref{sec:impurities} we analyze the parallel
momentum equation for the impurities and derive an equation that
governs their poloidal rearrangement. We show approximate solutions in
several limits and numerical solutions are also presented. In
Sec.~\ref{sec:flow} we explore the modification of the poloidal
impurity rotation due to the presence of large gradients, discussing all regimes of main-ion
collisionality.
Finally, the results are summarized and discussed in
Sec.~\ref{sec:conclusions}.

\section{Kinetics of main ions in the plateau regime}
\label{sec:ions}
The plasma is assumed to consist of hydrogenic ions
($i$) in the plateau regime, collisional
(Pfirsch-Schl\"{u}ter) impurities ($z$), and electrons ($e$).  The
 calculation does not depend on the collisionality regime
of the electrons.  The magnetic field
is represented as $\g B = I(\psi) \nabla \varphi + \nabla \varphi
\times \nabla \psi$, where $\varphi$ is the toroidal angle and $2\pi\psi$
is the poloidal flux.
Throughout this analysis we will use a poloidal angle coordinate
$\vt$ which is chosen so that $\g B \cdot \nabla \vt$ is a flux function.
 This coordinate
makes flux surface averages convenient to evaluate
($\lla Y \rra = (2\pi)^{-1}\int_0^{2\pi}Y d\vt$ for any quantity $Y$),
and this coordinate is equivalent to the $\vt$ used in \cite{perbif,fh1,fh2}.
We assume a model field magnitude $b^2=1 - 2\epsilon \cos \vt$ where
$b=B/\lla B^2 \rra^{1/2}$ and $\epsilon=r/R$ is the inverse aspect ratio.
We must assume $\epsilon \ll 1$ from the beginning of the analysis
in order for a plateau regime to exist.

The gyroaveraged ion
distribution function in the plateau regime is then given by \cite{pusztai}
$\bar{f}_i=f_{Mi} + \bar{f}_{i1}$ where
 \begin{equation}
 f_{Mi}=n_{i0}(\psi)\lp\frac{m_i}{2\pi T_i(\psi)}\rp^{3/2}\exp \left(-\frac{m_iv^2}{2 T_i(\psi)}\right)
 \end{equation}
is a stationary Maxwellian and a flux function,
\begin{equation}
\bar{f}_{i1}= -f_{Mi}\frac{e \Phi_1}{T_i}
+H_i-f_{Mi}\vpa\frac{I}{\Omega_i} \lp
\frac{p_i'}{p_i}+\frac{e \Phi_0'}{T_i} + \frac{y b^2 T_i'}{2 T_i}\rp,
\label{plateaudistr}
\end{equation}
$p_i=n_{i0}T_i$, $\Omega_i= e B/m_i$ is the ion cyclotron frequency, primes denote $d/d\psi$, $\Phi_0=\lla \Phi \rra$,
$\Phi_1 = \Phi - \Phi_0$,
\begin{equation}
  H_i=Q_i \frac{\hat{\nu}_i\sin \vt-\xpa\cos \vt}{\xpa^2+\hat{\nu}_i^2}\approx Q_i\left[\pi
    \delta  (\xpa)  \sin\vt-\frac{
      \cos\vt} {  \xpa  }\right],
\label{platsoli}
\end{equation}
$\hat{\nu}_i=\nu_i q R/v_i$ is the normalized collisionality,
$x=v/v_{i}$, $v_{i}= (2
T_i/m_i)^{1/2}$,
\begin{equation}
Q_i=f_{Mi} \frac{\epsilon v_i I T_i'}{4 \Omega_i T_i} \left[
(2 \xpa^2+\xpe^2)(2 x^2-5) + y b^2(2 \xpa^2-\xpe^2)\right],
\label{Qi}
\end{equation}
and $y$ is a velocity-independent coefficient which will be determined by
the requirement of ambipolarity.
In a pure plasma this requirement leads to $y=1$, but the presence of impurities will alter the value.
Also, $y$ must be a flux function so that
$\nabla\cdot( n_i \g V_i) = 0$.

\section{Impurity dynamics}
\label{sec:impurities}

The parallel momentum equation for the impurities is taken to be
\begin{equation}  0=-z n_z e
\nabla_\parallel \Phi - T_i \nabla_\parallel n_z +
R_{zi\parallel}
\label{pme}
\end{equation}
where $R_{zi\parallel}$ is the impurity-ion friction. The parallel viscosity of
the impurities has been neglected since it was shown in Ref \cite{perbif} to be smaller
than the pressure gradient if $\delta / z \hat{\nu}_{ii} \ll 1$, which
is usually the case in the tokamak edge.  As also shown in that paper,
the impurity temperature is then equilibrated with the bulk ion
temperature and is therefore constant over the flux surface.  
The poloidal electric field $-\nabla_{||} \Phi$ can be obtained from the
quasi-neutrality condition $z n_z=n_e-n_i$ using $n_e=(1+e\Phi_1/T_e)n_{e0}(\psi)$
 and using the  distribution function
(\ref{plateaudistr}) to calculate the ion density:
\begin{equation}
n_i=n_{i0}\left(1-\frac{e\Phi_1}{T_i} + \epsilon N_s \sin{\vt}\right),
\label{plateauni}\end{equation}
where
\begin{equation}
N_s=-\frac{\sqrt{\pi}
    v_i I T_i'}{4\Omega_i T_i} \left(  1 +b^2 y \right).
\end{equation}
The result is
\be\frac{ze
  \nabla_\parallel \Phi}{T_i}=\frac{T_0}{2 T_i n_0}
  \nabla_\parallel \left(z^2 n_z+  z n_{i0}\epsilon N_s\sin{\vt}\right), \ee
  where $2 n_0/T_0 \equiv n_{e0}/T_e+ n_{i0}/T_i$.
Equation (\ref{pme}) then becomes
\be
(1+\alpha n)\nabla_\parallel n
+\frac{\epsilon z T_0 n_{i0} n } {2 T_i n_0}
\nabla_\parallel\lp N_s \sin{\vt}\rp
=\frac{R_{zi\parallel}}{\langle n_z\rangle T_i}
\label{pme1}
\ee where $n = n_z / \langle n_z \rangle$ is the normalized impurity density
and $\alpha\equiv \lla n_z \rra z^2 T_0/(2 n_0 T_i)$.
In the rest of the analysis we will order $\alpha \sim 1$, which is
equivalent (for $T_e \sim T_i$) to the ordering $z_{eff}-1 \sim 1$.

Next, the ion-impurity collision operator $C_{iz}$ is inserted
in $R_{zi\parallel} = -m_i \int d^3v\, v_{||}C_{iz}$ to write
\be
R_{zi\parallel}=-\int d^3v\, m_i v_\parallel \nu_{iz}\left({\cal
  L}(f_i-f_{i0})+\frac{m_i v_\parallel}{T_i}V_{z\parallel} f_{i0}\right)
\ee
where
\be
{\cal L} =\frac{2 v_\parallel}{v^2 B} \frac{\partial }{\partial
\lambda}  \lambda v_\parallel \frac{\partial }{\partial \lambda}
\ee
is the Lorentz pitch-angle scattering operator, $\lambda =v_\perp^2/(B v^2)$,
$ \nu_{iz}=3 \pi^{1/2} / (4 \tau_{iz} x^3)$, and
$\tau_{iz}=3(2\pi T_i)^{3/2} \epsilon_0^2 m_i^{1/2}/(n_z z^2 e^4)$ is the ion-impurity collision time.
To ensure $\nabla\cdot (n_z \g V_z) = 0$, the parallel impurity flow velocity
must have the form \cite{perbif}
\be
 V_{z\parallel}=-\frac{I \Phi_0'}{B}+\frac{K_z(\psi)
B}{n_z}
\label{formOfFlow}
\ee
where $K_z(\psi)$ is proportional to the poloidal velocity.
Using the main-ion distribution function (\ref{plateaudistr})
we then obtain
\be
R_{zi\parallel}= -\frac{I}{\Omega_i \tau_{iz}}
\left( p_i' + \frac{y b^2 n_{i0} T_i'}{2} \right)
 -\frac{m_i n_{i0} K_z B}{\tau_{iz} n_z} + Q_r,
\label{friction}
\ee
where
\be
Q_r=m_i \int d^3 v \nu_{iz} v_\parallel Q_i \left[\pi \delta(x_\parallel)\sin{\vt}-\frac{x_\parallel }{x_\parallel^2+\hat{\nu}_i^2}\cos{\vt}\right].
\label{qr}
\ee
For $\hat\nu_i \to 0$ the integration results in
\be Q_r= 3\frac{\epsilon n_{i0}
  I T_i^\prime}{\tau_{iz}\Omega_i}\cos{\vt}.
\label{friction1}
\ee
To rewrite Eq.~(\ref{pme1}) in dimensionless form we introduce
the ratio of the temperature and pressure scale lengths $\eta = p_i T_i' / (T_i p_i')$,
\be
g=-\frac{m_i I p_i'}{e T_i \tau_{iz} n_z \g B \cdot \nabla \vt},
\ee
and
\be
\tau_*=\frac{\sqrt{\pi} z T_0 \tau_{iz} n_z v_i \g B \cdot \nabla\vt}{8 T_i n_0 B_0},
\ee
where $B_0=\lla B^2\rra^{1/2}$.
Notice that $g$, $\tau_*$, and $(\tau_{iz}n_z)$ are $\vt$-independent, and the formal magnitude of
$\tau_*\sim (z \hat\nu_i)^{-1}$ has not yet been fixed.
%
Equation (\ref{pme1}) now becomes
\begin{eqnarray} (1+\alpha n) \frac{\partial n}{\partial \vt}
=g\left\{ n +\frac{\eta y n b^2}{2}
- \epsilon \eta \left[3+(1+y)\tau_* \right] n \cos{\vt}
+K_z \frac{n_{i0} e B^2}{\lla n_z \rra I p_i'}
 \right\}
\label{pme2}
\end{eqnarray}
where we have used $\partial (N_s \sin\vt)/ \partial\vt \approx \lla N_s \rra \cos\vt$.
(Other terms of order $\epsilon$ have already been discarded in deriving the distribution function
(\ref{plateaudistr}).)
Integrating Eq
(\ref{pme2}) over $\vt$ yields a solubility constraint
which can be used to determine the poloidal impurity rotation,
\be
K_z = \frac{ \lla n_z \rra I p_i'}{n_{i0} e \lla B^2 \rra}
\left\{ -1
- \frac{\eta y}{2} \lla n b^2 \rra
+ \left[3+(1+y)\tau_* \right] \epsilon \eta \lla n \cos\vt \rra
\right\},
\label{kz}
\ee
and Eq (\ref{pme2})  becomes
\begin{eqnarray} (1+\alpha n) \frac{\partial n}{\partial \vt}
=g\left[  n - b^2
+ \frac{ \eta y b^2}{2} \lp n - \lla n b^2 \rra \rp
\right.
\hspace{5cm}\nonumber\\
\left. +\left[3+(1+y)\tau_* \right] \epsilon \eta  \lp b^2 \lla n \cos\vt\rra - n \cos\vt \rp
\right].
\label{pme3}
\end{eqnarray}
The $\cos\vt$ terms above can be significant
despite being proportional to $\epsilon$, for the other drive
in the equation is the $\vt$-variation in $b$, which is also $\ord(\epsilon)$.

To make further progress we will calculate the coefficient $y$
by requiring ambipolarity. Due to the smallness of the electron mass, the ambipolarity condition is approximately
$\Gamma_i=-z \Gamma_z$. As in the conventional plateau-regime calculation for a pure plasma,
the main-ion flux is
\begin{equation}
  \Gamma_i\equiv\langle \g \Gamma_i\cdot \nabla \psi\rangle
  = \frac{\sqrt{\pi} \epsilon^2 v_i^3 I^2 (\g B \cdot\nabla\vt) n_{i0} T_i'}
  {8 \Omega_{i0}^2 B_0 T_i} (y-1)
\end{equation}
where $\Omega_{i0}=e B_0/m_i$.
The impurity flux is driven by the impurity-ion parallel friction force
\be
\Gamma_z\equiv\lla \g \Gamma_z\cdot \nabla \psi \rra=\lla \frac{I
  R_{zi\parallel}}{z e B}\rra,
\ee
where  $R_{zi\parallel}$ is given by Eq.~(\ref{friction}) with $K_z$
from (\ref{kz}). We find
\begin{eqnarray}
\Gamma_z= \frac{m_i I^2 \langle n_z\rangle p_i'}{z e^2 \tau_{iz}n_z
  \langle
  B^2\rangle}\left(
1 - \lla\frac{n}{b^2}\rra
+\frac{\eta y}{2} \left[\langle n b^2\rangle-1 \right] \right.
\hspace{5cm}\nonumber\\
\left.
+\epsilon\eta\left\{ 3 \lla \frac{n  \cos{\vt}}{b^2}\rra- \left[ 3+ (1+y)\tau_* \right] \lla n \cos{\vt}\rra \right\} \right).
\end{eqnarray}
The condition for ambipolarity then gives
\be
y=\frac
{z \epsilon^2 \tau_* \alpha^{-1} + \eta^{-1}( \lla n/b^2 \rra -1 ) - 3 \epsilon \lla n b^{-2} \cos\vt \rra + (3+\tau_*)\epsilon \lla n \cos\vt \rra }
{ z \epsilon^2 \tau_* \alpha^{-1} + 2^{-1}(\lla n b^2 \rra-1) - \epsilon \tau_* \lla n \cos\vt \rra}.
\label{y}
\ee
The pure plasma limit $y=1$ is recovered as $\alpha \to 0$.

The system (\ref{pme3}) and (\ref{y}) describes the poloidal rearrangement of the
impurities. While (\ref{pme3}) is similar to the equations found if the main ions
are in the banana~\cite{perbif,fh1} or \ps\ regimes~\cite{fh2}, (\ref{pme3}) has several different terms,
and also the radial scale length entering $g$ is different (i.e. only the pressure scale length
appears, rather than a combination of the pressure and temperature scale lengths.)
 As in Refs.\cite{perbif,fh1,fh2}, $g$ measures the
steepness of the bulk ion pressure profile.  In conventional
neoclassical theory $g$ is assumed to be small, which implies that the
friction force is smaller than the parallel pressure gradient.
We next examine how the integro-differential equation (\ref{pme3})
can be solved analytically in a number of limits.

\paragraph{Weak density variation.}

If $n-1 \sim \ord(\epsilon)$ then we can expand $n=1+n_c \cos{\vt}+ n_s \sin{\vt}+ O(\epsilon^2)$
with $n_s$ and $n_c$ both $\sim\ord(\epsilon)$.
The solution of Eq (\ref{pme3}) is then found to be
\begin{eqnarray}
n_s & =&  \epsilon g (1+\alpha) \frac{2-\eta [3+(1+y)\tau_*]}{(1+\alpha)^2+
g^2 (1+\eta y/2)^2},  \\
n_c &=&  -\epsilon g^2 (1+\eta y/2) \frac{2-\eta [3+(1+y)\tau_*]}{(1+\alpha)^2+
g^2 (1+\eta y/2)^2}.
\label{nc}
\end{eqnarray}
It can be noted from these expressions that as $p_i'$ becomes
larger, the impurities first develop an up-down asymmetry and then an in-out asymmetry.
This same behaviour is found in the banana and
\ps\ regimes. However, in the plateau regime the asymmetry is proportional to the new factor
$2-\eta [3+(1+y)\tau_*]$, which means that the sign of the asymmetry can be
changed depending on the magnitude of $\eta$, $\tau_*$ and $y$. If $\eta> 2/[3+(1+y)\tau_*]$, the impurities will be pushed to the outside of the
flux surface. This result is different from the analogous $n-1 \sim \ord(\epsilon) \ll 1$
limits when the main ions are in the banana or \ps\ regimes.  In these cases, in the absence of rotation, the
impurities were pushed to the inside, regardless of the ratio of the
pressure and temperature gradients.

\paragraph{Large gradients.}

In the $g \gg 1$ limit, corresponding to a large pressure gradient,
we can expand (\ref{pme3}) in $g^{-1}$. To lowest order,
the right-hand side of (\ref{pme3}) must vanish, giving $n \approx \tilde{n} / \lla \tilde{n} \rra$
where
\be
\tilde{n} =
\frac{b^2}{1+(\eta y/2) b^2-\epsilon \eta [3+(1+y)\tau_*]\cos{\vt}}.
\label{largeglimit}
\ee
In this case there is only in-out asymmetry. Expanding in $\epsilon$ then gives
\be
n= 1+2\epsilon (S-1)\cos\vt
\label{largeglimit2}
\ee
where $S=\eta\left[3+y+(1+y)\tau_*\right]/(2+\eta y)$.
(This same result can also be obtained by a $g \gg 1$ expansion of (\ref{nc}).)
 The impurity density evidently may be
higher at either the inboard side $(S<1)$ or outboard side $(S>1)$.
This finding too differs from the corresponding $g \gg 1$ limits
when the main ions are in the banana or \ps\ regime.
In these cases, the impurity density is always
greater at the inboard side (even when there is significant rotation).

\paragraph{Numerical solution.}

\begin{figure}[htbp]
\begin{center}
 \includegraphics[width=0.9\textwidth]{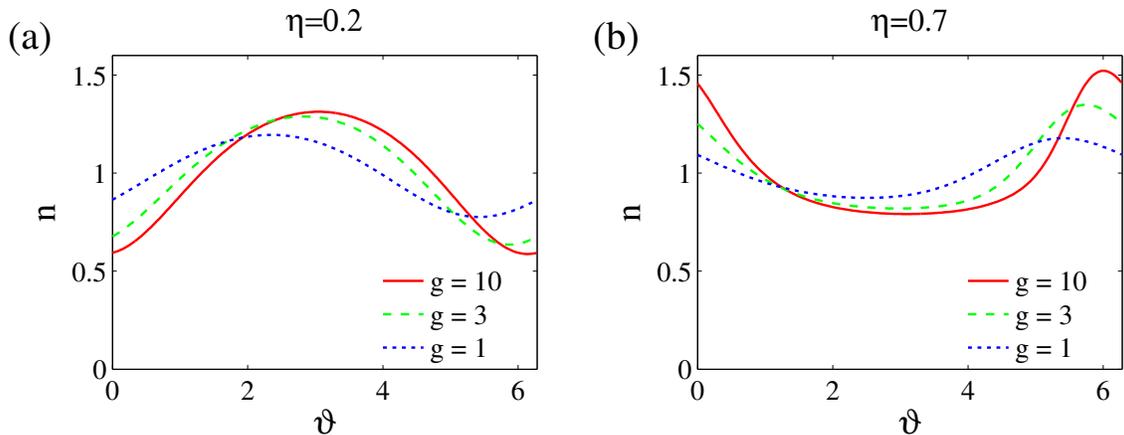}
\caption{(Color online) Normalized impurity density as function of
  poloidal angle, calculated by numerical solution of (\ref{pme3}) and (\ref{y}). The parameters used are $\epsilon=0.3$, $\tau_\ast=0.5$, $z=5$, and $\alpha=0.25$.
   Other values of $\alpha$ from 0 to 1 produce nearly indistinguishable results.}
\label{noftheta}
\end{center}
\end{figure}

\begin{figure}[htbp]
\begin{center}
\includegraphics[width=0.9\textwidth]{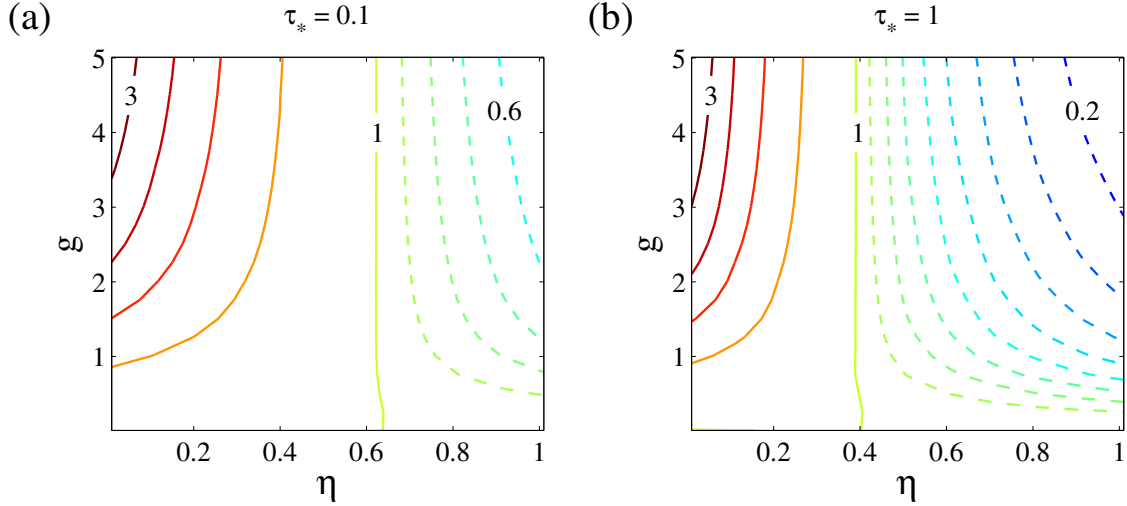}
\caption{(Color online) Contours of the in-out asymmetry $A$, decreasing monotonically with $\eta$, for (a) $\tau_*=0.1$ and (b) $\tau_* = 1.0$.
The other parameters are $\epsilon=0.3$, $z=5$, and $\alpha=0.25$.
Results for $\alpha=0$ are nearly indistinguishable. Solid contours run from $A=3$ to 1 in steps of 0.5.
Dashed contours decrease from $A=0.9$ to 0.6 (a) or 0.2 (b) in steps of 0.1.}
\label{asymmFig}
\end{center}
\end{figure}

For $\alpha \ll 1$, equation (\ref{pme3}) may be solved numerically with the following iterative procedure.
A small number (5-10) of poloidal Fourier modes are considered.
An initial guess for $n(\vt)$ is used to compute $y$ and the nonlinear term $\alpha n \; \partial n/\partial \vt$.
An improved $n(\vt)$ is then calculated using (\ref{pme3}), and the process is repeated until convergence is achieved.
Typical results are shown in Figure \ref{noftheta}.  Figure \ref{asymmFig} shows the in-out asymmetry factor
\be
A=\frac{n(\vt=\pi)}{n(\vt=0)}
\ee
over a wide range of parameters.

\begin{figure}[htbp]
\begin{center}
 \includegraphics[width=0.53\textwidth]{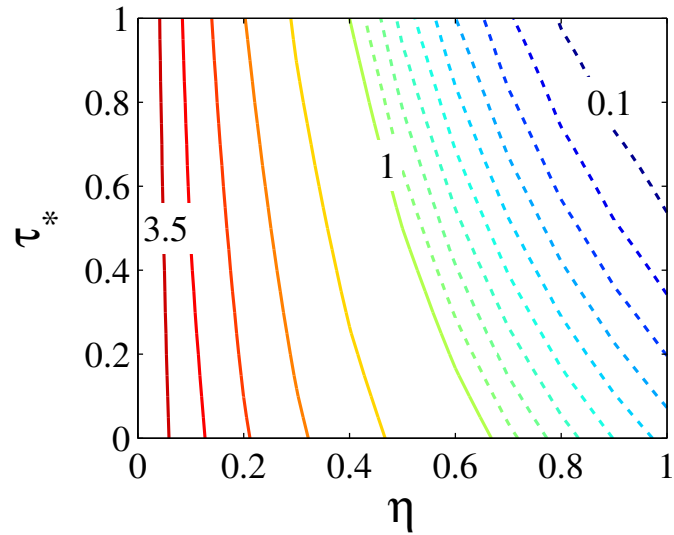}
\caption{(Color online) Contours of the in-out asymmetry factor $A$
in the $g\gg 1$ and trace impurity ($y\to 1$) limit and with $\epsilon=0.3$.
Solid contours range from 3.5 to 1 with spacing of 0.5,
and dashed contours range from 0.9 to 0.1 with spacing of 0.1.
}
\label{largegFig}
\end{center}
\end{figure}

Figure (\ref{largegFig}) shows the in-out asymmetry $A$
for  $\epsilon=0.3$, $g=10$, and the trace impurity limit $y\to 1$.  A nearly
identical plot can be generated using the $g \gg 1$ expressions (\ref{largeglimit}) or (\ref{largeglimit2}),
although the precise value of $A$ in the $A<1$ region is
somewhat different due to the fact that $\epsilon=0.3$
is not much smaller than one.

\section{Poloidal impurity rotation}
\label{sec:flow}
If the impurity density varies on a flux surface, the impurity poloidal
rotation will be different from the one derived in conventional
neoclassical theory. Using (\ref{formOfFlow}) and (\ref{kz}), we can write
\begin{equation}
  V_{z\vt}^{pl}=\frac{B_\vt K_z }{n_z}
=- X \frac{IB_\vt}{ n e \lla B^2 \rra}
\left[ \frac{T_i}{n_{i0}} \frac{d n_{i0}}{d\psi} + \frac{3}{2} \frac{d T_i}{d\psi} \right]
,
\label{poloidalFlow1}
\end{equation}
where
\begin{eqnarray}
X = \left( 1 + \frac{\eta}{2}\right) ^{-1}
\left\{1 + \frac{\eta y}{2} \lla nb^2 \rra - \left[ 3+ (1+y)\tau_* \right] \epsilon \eta \lla n \cos\vt\rra \right\}
\end{eqnarray}
is constant on a flux surface.  The definition of $X$ was chosen above so that
in the trace impurity limit ($\alpha\to 0$, $y\to 1$) and if $n_z$ is also uniform on a flux surface (i.e. $g \to 0$),
then $X \to 1$.  This limit reproduces the conventional neoclassical result \cite{HS, cattosimakov}.

\begin{figure}[htbp]
\begin{center}
 \includegraphics[width=0.48\textwidth]{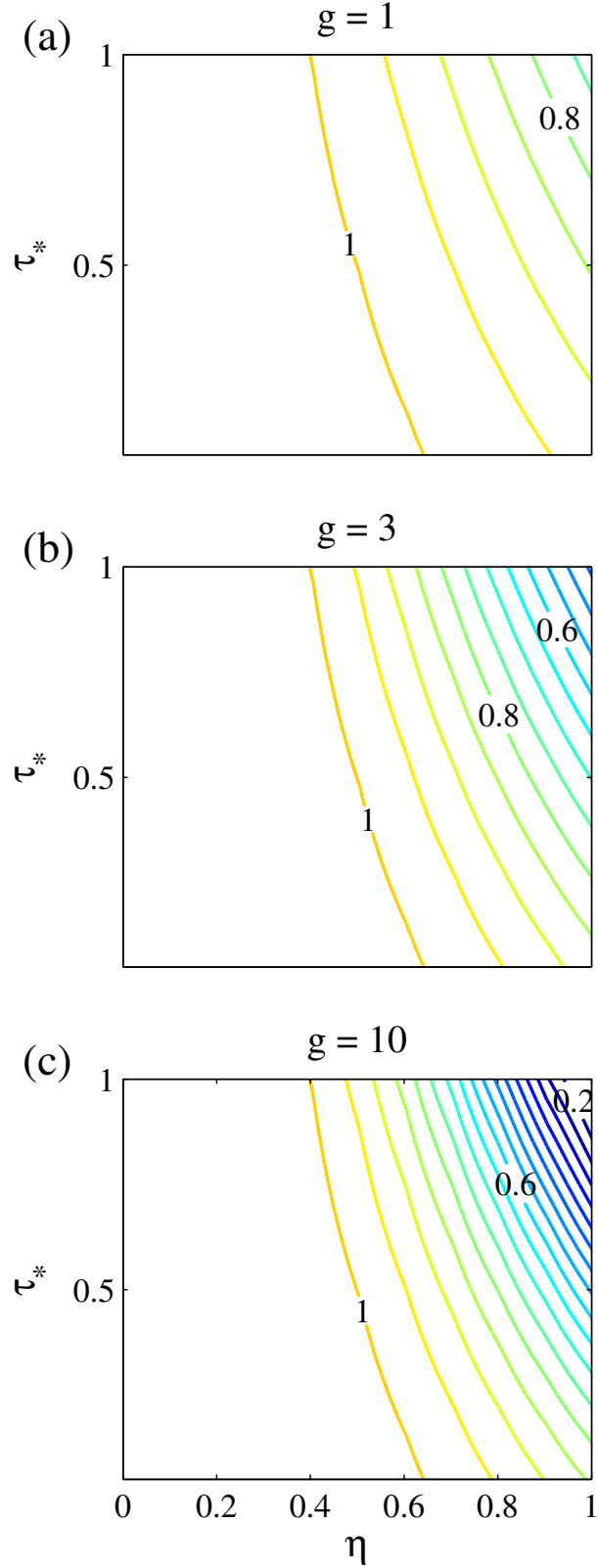}
 \caption{(Color online) The factor $X$ which scales the poloidal impurity flow
 in the plateau regime, calculated for $\alpha=0$. The horizontal axis is the same for all plots.  Contour spacing is 0.05.}
 \label{Xalpha0Fig}
\end{center}
\end{figure}

\begin{figure}[htbp]
\begin{center}
 \includegraphics[width=0.48\textwidth]{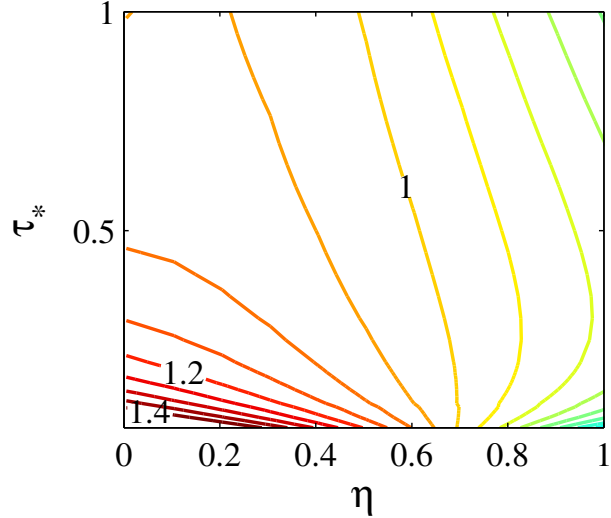}
 \caption{(Color online) The factor $X$ which scales the poloidal impurity flow
 in the plateau regime for $\alpha=0.25$ and $g=1$. Contour spacing is 0.05.}
 \label{Xalpha025Fig}
\end{center}
\end{figure}

Figure \ref{Xalpha0Fig} shows the scale factor $X$ for
various values of $\eta$, $\tau_*$, and $g$.  The figure was calculated
using $\epsilon=0.3$ and $\alpha \to 0$.  It is evident that when $g>1$, the poloidal
flow can be significantly suppressed compared to the conventional neoclassical result
if $\eta$ and $\tau_*$ approach one.
The situation is only slightly different when the relative impurity strength $\alpha$
is nonzero, as shown in Figure \ref{Xalpha025Fig}. This figure is equivalent to Figure \ref{Xalpha0Fig}.a
but with $\alpha$ raised to 0.25 and $z=5$.  When $\tau_* \ll 1$, the flow now
becomes slightly enhanced compared to the conventional
neoclassical result.

When the main ions are in the banana regime, the poloidal impurity flow
can be calculated using the $K_z$ derived in Refs. \cite{perbif} and \cite{fh1}. The result is
\be V_{z\vt}^{ban}= \frac{K_z B_\vt}{n_z} =
\frac{B_\vt}{n} u \lp \lla n b^2 \rra +
\frac{1}{\gamma^{ban}}  \rp,
\label{poloidalFlowBanana}
\ee
where $\gamma^{ban} =e L_{\perp,ban} \lla B^2 \rra u/T_i$
and $ L_{\perp,ban}^{-1} = -I \lp p_i^\prime/p_i-(3/2)T_i^\prime/T_i\rp$.
In the limit of trace impurities and large aspect ratio,
      \be u =  -0.33 f_c \frac{I}{e \lla B^2 \rra}\frac{d
    T_i}{d \psi},
    \label{ualfaliten}
    \ee
and
\be f_c \equiv \frac{3 \lla B^2 \rra}{4}
        \int_0^{\lambda_c}
        \frac{\lambda d \lambda}{\lla n\sqrt{1-\lambda B}\rra}
        \label{fc}
\ee is the effective fraction of circulating particles.
Therefore, in this limit,
\begin{equation}
  V_{z\vt}^{ban}=
- \frac{I B_\vt}{ n e \lla B^2 \rra}
\left[ \frac{T_i}{n_{i0}} \frac{d n_{i0}}{d\psi} + \left(-\frac{1}{2} + 0.33 f_c \lla nb^2 \rra \right) \frac{d T_i}{d\psi} \right]
,
\label{poloidalFlowBanana1}
\end{equation}
The
expression for $u$ in various other limits (arbitrary aspect
ratio and high level of impurities) is more complicated and is
given in Ref.~\cite{fh1}.

When the impurity density is nearly constant on a flux surface,
(\ref{fc}) gives the conventional result $f_c \approx 1-1.46\sqrt{\epsilon}$. For insight into
 how $f_c$ is modified when the impurity density varies significantly on a flux surface,
 consider the limit $n=\delta(\vt-\pi)$ in which the impurities are strongly peaked on the inboard
 midplane.  Then $\lla n \sqrt{1-\lambda B} \rra = \sqrt{1-\lambda B_{max}}$ so
 $f_c \approx 1-2\epsilon$.

Similarly, when the main ions are in the \ps\ regime, the poloidal impurity flow $V_{z\vt}^{PS}$
can be calculated using the $K_z$ derived in equation (26) of Ref. \cite{fh2}. For trace impurities, $V_{z\vt}^{PS}$
is found to be
\begin{equation}
  V_{z\vt}^{PS}=
- \frac{I B_\vt}{ n e \lla B^2 \rra}
\left[ \frac{T_i}{n_{i0}} \frac{d n_{i0}}{d\psi} + 2.8 \lla nb^2 \rra \frac{d T_i}{d\psi} \right]
,
\label{poloidalFlowPS}
\end{equation}

It was found in Refs. \cite{perbif} and \cite{fh1} that when the main ions are in the
banana or \ps\ regimes, the impurities tend to accumulate on the high field side, so $\lla nb^2 \rra >1$.
In both regimes, this change decreases the signed $V_{z\vt}$, shifting the poloidal impurity flow
in the direction of the electron diamagnetic velocity relative to the conventional
neoclassical prediction.  We can model the impurity density variation
as $n=1-(A-1)(A+1)^{-1}\cos\vt$, implying $\lla nb^2 \rra = 1+\epsilon (A-1)/(A+1)$.
As $A$ increases above one, $\lla nb^2 \rra$ increases from one to $1+\epsilon$.
For the banana regime, this increase in $\lla nb^2\rra$
and the aforementioned increase in $f_c$ both lead to a decrease in the signed $V_{z\vartheta}$,
with the $O(\sqrt{\epsilon})$ increase in $f_c$ being the larger of the two effects.

Note that in the method used in this section, the impurity
pressure gradient $p_z'$ does not appear in the formulae for the poloidal impurity flow for
any collisionality regime (as it does in, for example, equation
(15) of \cite{cattosimakov}).  In the conventional neoclassical formulae, the
$p_z'$ term is proportional to $1/z$, so the term is
formally small in our ordering.  The absence of the $p_z'$ term
is related to the fact that the impurity diamagnetic flow was
dropped in Eq.~(\ref{formOfFlow}) in order to make the analysis tractable.

\section{Conclusions and discussion}
\label{sec:conclusions}
In this paper we have investigated the poloidal rearrangement of impurities
in the presence of large gradients for the case of background ions
in the plateau collisionality regime.
The calculation shows that when the temperature scale length is large compared to the density scale length
(such that $\eta<0.4-0.6$), the impurities accumulate on the inboard side,
whereas they accumulate on the outboard side in the opposite case.
In standard tokamak operating regimes, $\eta<0.5$, so impurity accumulation
at the inboard side is more likely.  However, $\eta$ can be larger than 0.5
in the I-mode regime of Alcator C-Mod \cite{Imode}, so the strong $\eta$-dependence
of $A$ predicted by the theory
may be experimentally testable.
  (Impurity asymmetry in I-mode has not been measured
as of this writing.)

One way in which the present calculation could be extended would be to account for
the large radial electric field $E_r$ which arises in the pedestal.
It is found experimentally that the radial
electric field in the pedestal
can be large enough to make the ${\g E}\times {\g B}$ drift
 comparable to
$(B_{\theta}/B) v_i$, and it was recently shown in \cite{pusztai} that under
these conditions, the plateau-regime ion distribution function can
deviate from Eqs.~(\ref{platsoli}-\ref{Qi}).  Although it would be
desirable to include this effect in the present calculation of
impurity asymmetry, doing so is not straightforward, for the following reason.
Terms in the ion distribution function of order
$(\rho_{\theta}/a)\epsilon f_{Mi}$ affect the impurity asymmetry
calculation to leading order, as demonstrated by the term with a factor of 3 in
Eq.~(\ref{pme3}), which arises due to the $Q_r$ term in
Eq.~(\ref{friction}).  However, the ion distribution function in
\cite{pusztai} is only determined to order $(\rho_{\theta}/a)\epsilon^0
f_{Mi}$, and to consistently determine all $O(\epsilon)$ corrections,
the ion distribution function would need to be found using the full
linearized Fokker-Planck collision operator rather than a Krook or
pitch-angle scattering model operator.

In all regimes of main-ion collisionality,
the poloidal rearrangement of impurities results in changes to the
the poloidal impurity flow.  These modifications to the flow are of interest because
when the main ion collisionality is in
the plateau or banana regimes in Alcator C-Mod,
impurity velocity in the pedestal is measured to be greater in the electron
diamagnetic direction than conventional neoclassical
theory predicts \cite{marr, newcmod}.
When the ions are in the plateau regime, the calculation in this paper
shows the impurity flow should be multiplied by the factor $X$ relative to the conventional
neoclassical prediction (in which the flow is always in the electron diamagnetic direction.)
To explain the observed flows, then, $X$ must be $>1$, which can occur for $\tau_* \ll 1$ (as in Figure \ref{Xalpha025Fig}.)
When the ions are in the \ps\ regime,
we find the poloidal impurity flow is indeed increased in the
direction of the electron diamagnetic velocity due to the
increase in $\lla n b^2\rra$ above one.
For banana-regime ions,
the flow is shifted in the same direction due to both the increase in  $\lla n b^2\rra$
and also due to the increase in $f_c$.
However, the shift in the flow is also proportional to the small numerical factor 0.33 in (\ref{poloidalFlowBanana1}),
so this effect is likely insufficient to explain the observed discrepancy between the measured
and predicted flows.
A different calculation, including the $E_r$ effects discussed above but neglecting
the impurity asymmetry, is discussed in Ref. \cite{newcmod}; this calculation
can also explain some but not all of the discrepancy.  In future work, it may be possible to consistently
account for both the $E_r$ and impurity asymmetry effects simultaneously to achieve better agreement between the calculated and
observed flows.

\section*{Acknowledgements}
The authors gratefully acknowledge helpful conversations with Istvan
Pusztai, Peter J Catto, and Per Helander.  Two of the authors (M L and
D G) acknowledge the hospitality of Chalmers University of Technology,
where part of this research was carried out. This work was funded by the
European Communities under Association Contract between EURATOM and
{\em Vetenskapsr{\aa}det} and US Department of Energy Grant No
DE-FG02-91ER-54109. The views and opinions expressed herein do not
necessarily reflect those of the European Commission.
\bibliographystyle{unsrt}

\begin{thebibliography}{11}
\bibitem{perbif} P. Helander, Phys. Plasmas {\bf 5}, 3999 (1998); The
right-hand side of the last equation on p 4002 in this paper should be
multiplied by a factor $\epsilon / q$, and the right-hand side of the
previous equation (the expression for $n_c$) should be multiplied by
$-1$.
\end{thebibliography}

\end{document}